\documentclass[10pt,a4paper]{article}
\usepackage{graphpap,epsfig,amssymb,graphicx,textcomp,amsmath,xcolor}

\usepackage{multirow}
\usepackage{multicol}
\usepackage[export]{adjustbox}
\usepackage{hyperref}

\begin{document}

\begin{center}

{\LARGE {\bf Strange relaxation and metastable behaviours of the Ising ferromagnetic thick cubic shell}}

\vskip 0.5cm

{\large {\it Ishita Tikader$^1$ and Muktish Acharyya$^{2,*}$}} \\
{\it Department of Physics, Presidency University,}\\
{\it 86/1 College Street, Kolkata-700073, INDIA}\\
{$^1$E-mail:ishita.rs@presiuniv.ac.in}\\
{$^2$E-mail:muktish.physics@presiuniv.ac.in}

\end{center}

\vskip 1cm

\noindent {\bf Abstract:} 
We have studied the equilibrium and nonequilibrium behaviours of the Ising ferromagnetic thick cubic shell by Monte Carlo simulation. Our goal is to find the dependence of the responses on the thickness of the shell. In the equilibrium results, we found that the pseudo-critical temperature of ferro-para phase transition of thick cubic shell increases with the increase of the thickness following a hyperbolic tangent relation. In the nonequilibrium studies, the relaxation time has been found to decrease with the increase of the thickness of the cubic shell. Here three different regimes are found, namely rapid fall, plateau and linear region. The metastable behaviour has been studied also as another kind of non-equilibrium response. The metastable lifetime has been studied as function of the thickness of the cubic shell. A non-monotonic variation of metastable lifetime with the thickness of the shell is observed. A specified thickness for longest-lived metastability has been identified.
\vskip 3cm



\vskip 2cm

\noindent $^*$ Corresponding author
\newpage

\section{\bf {Introduction:}}

\textcolor{blue}{The Ising model, even after a century since its inception, remains a central and fascinating field in the research of magnetism. The equilibrium phase transition\cite{onsager,stanley,ito-suzuki} in Ising ferromagnet has been historically significant in the development of equilibrium statistical physics. Beyond the classic studies on regular geometric lattices, investigations of the ferro–para phase transition on the fractal lattices have uncovered\cite{fractal1,fractal2, fractal3} a range of intriguing phenomena. The structure and geometry play a crucial role on the critical behaviour of Ising ferromagnetic systems\cite{ishita-review}.}

\textcolor{blue}{The nonequilibrium behaviour of the Ising ferromagnets have also garnered substantial attention from the researchers . Among these, the relaxation dynamics stand out as a particularly intriguing aspect. Over the past few decades, numerous studies have focused on various aspects of relaxation, including the decay of metastable states, domain growth, and phase-ordering kinetics — topics that have come to define a major portion of the literature in nonequilibrium statistical mechanics.}

\textcolor{blue}{Historically, the relaxation dynamics was first studied by Suzuki and Kubo\cite{suzuki} in a Glauber kinetic Ising ferromagnet using mean-field approximation. The linearized mean-field dynamical equation (near the critical temperature) can give rise to the exponential relaxation of magnetisation along with  the critical slowing down. Further investigations\cite{stauffer1}, particularly through extensive Monte Carlo (MC) simulations, have explored relaxation behaviour both near and away from the critical temperature.}

The evidence of non-exponential relaxation can be found experimentally in magnetic nanoparticle \cite{gresits} for hyperthermia and manganese single-molecule magnet\cite{yamaguchi}. \textcolor{blue}{The Monte Carlo studies have also revealed stretched exponential relaxation in the large-scale Ising systems below the critical temperature\cite{stauffer2}.}
Extremely slow relaxation has been observed\cite{guo} near the first order transition in the three-dimensional Ising ferromagnet. Recently, the effects of boundary conditions, geometry, and dynamical rules have been studied\cite{ishita1} in a two-dimensional Ising ferromagnet by employing MC simulation.

Ising ferromagnetic system has been widely used over the past few decades to study nucleation \cite{becker, grant,vehkamaki,binder1}, domain growth kinetics \cite{puri} and metastable behaviour. The magnetic field dependence of metastable lifetime has been investigated 
\cite{rikvold,acharyya} extensively in the Ising ferromagnet by using MC simulation. The heat-assisted magnetisation reversal in ultra-thin ferromagnetic film has been studied \cite{deskins} using kinetic Monte Carlo simulation, leading to enhanced nucleation. The magnetisation reversal in general spin-{\it{\bf s}} Ising ferromagnetic film has been studied\cite{mapre,moumita-review} recently by employing MC simulation to find the role of anisotropy. Interestingly, an effective two-dimensional behaviour of metastability, has recently been reported\cite{ishita2} for a thin Ising ferromagnetic cubic shell 
\textcolor{blue}{despite of its three
dimensional Euclidean geometry.} These studies hold significance beyond just pedagogical interest. \textcolor{blue}{Real-world applications such as magnetic data storage and recording technologies\cite{storage,vogel}}
require an understanding of the dynamical aspects of ferromagnetic samples, which can be visualized through MC simulations.

In the above mentioned survey, we have not found any such studies of equilibrium and nonequilibrium behaviours of the thick 
Ising ferromagnetic shell where the thickness may influence on the behaviours of the ferromagnet. In this manuscript we report
the results of our MC investigations of the equilibrium and nonequilibrium responses of a thick Ising ferromagnetic cubic
shell and studied the physical quantities as function of the thickness of the shell. The paper is organised as follows:
the next section introduces the model and the simulation methodology. Section-3 contains the simulational results with data
and figures. The papers ends with a summary in Section-4.

\section {\bf {Model and Simulation method:}}
The classical Ising model is described on a simple cubic shell. The total energy of the system with uniform nearest-neighbour interaction, placed in a uniform external magnetic field $h_{\rm ext}$ is determined by the following Hamiltonian, 
\begin{equation}
\label{hamiltonian}
H=-J\sum_{<i,j>}S_iS_j - h_{ext} \sum_{i}S_i
\end{equation}
where spin at $i^{th}$ site $S_i$ can take two discrete values +1 and -1 only i.e., $S_i \in \{-1, +1\}$; $\forall~i$. $J(>0)$ denotes the uniform ferromagnetic coupling between nearest-neighbour spins.

\begin{figure}[h!tpb]
 \centering
  \includegraphics[angle=0, width=0.70\textwidth]{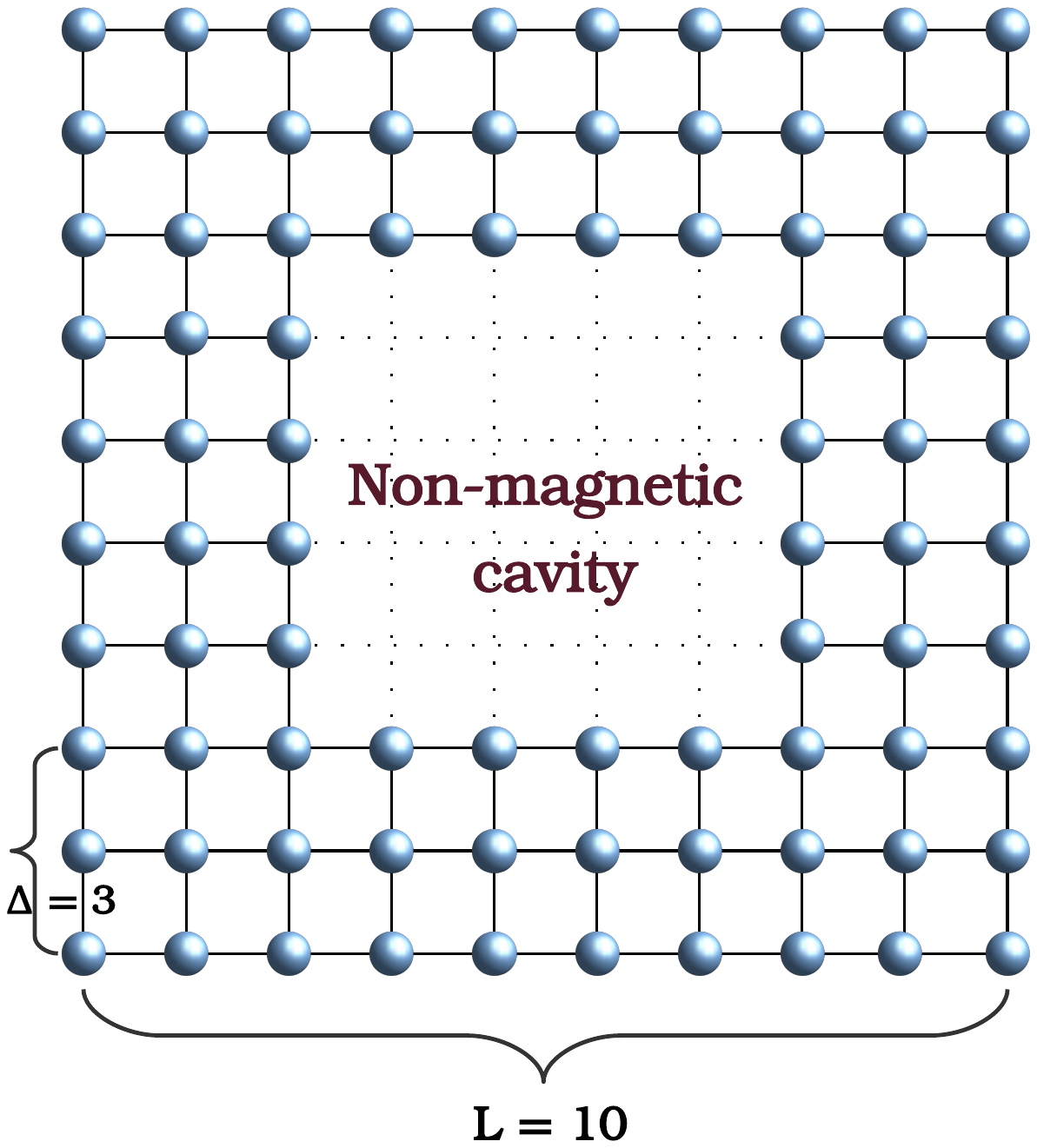}
  \caption{A schematic cross section of a cubic shell of size $L$ and thickness $\Delta$. Total lattice sites $N_s = L^3 - (L-2\Delta)^3$. Here we have demonstrated a cubic shell of $L=10$ having thickness $\Delta = 3$. Inside the shell there exists non-magnetic hollow space.}
  \label{fig:spinstructure}
\end{figure}

 We have considered the Cubic shell of size $L$, having thickness $\Delta$ with open boundary conditions in all three directions. The system contains the total sites of $N_s= L^3 - (L-2\Delta)^3$. A schematic cross-sectional diagram of the cubic shell is provided in Figure \ref{fig:spinstructure}. Within the thick outer shell there exists a hollow non-magnetic cavity.

The spins in the kinetic Ising model interact with a large heat bath at a constant temperature that induces spontaneous random spin-flips (only one spin at once). Monte Carlo simulation with Metropolis single spin algorithm is employed here to study the dynamics of kinetic Ising model. Starting from our initial configuration, we have chosen a site in random manner (say $i$). At any fixed temperature ($T$) and external field ($h_{ext}$) the spin-flip ($S_i$) at randomly chosen site will be accepted based on transition probability determined by Metropolis rule\cite{binder2}.
\begin{subequations}
    \begin{equation}\label{prob}
       P(S_i \rightarrow  -S_i) = Min \Biggl[ 1, \exp \biggl(-\frac{\Delta E_i}{k_BT} \biggr) \Biggr]  
    \end{equation}
\rm{where $\Delta E_i$ is the change in energy (in unit of $J$) caused by the proposed change in spin configuration at $i^{th}$ site.}
    \begin{equation}\label{local_energy}
      \Delta E_i =  2S_i \Biggl[ J\sum_{NN}S_j + h_{ext} \Biggr]
    \end{equation}
\end{subequations}
$k_B$ is the Boltzmann constant and $T$ is the temperature of the system measured in the unit of $J/k_B$. Here, $J =1$ and $k_B = 1$ are chosen conventionally for simplicity. According to the protocol, if probability $P(S_i \rightarrow  -S_i)= 1 $ then the spin must flip. If $P(S_i \rightarrow  -S_i) < 1 $, then the spin will flip with probability $P$. An uniformly distributed (in the range $ [0,1] $) random number `$r$' is generated. Now, the proposed spin-flip is accepted if the condition $r\le P(S_i \rightarrow  -S_i)$ is satisfied. Otherwise, the proposal of spin-flip will be rejected and the spin will retain its original state. This randomly selected $N_s$ number of updates constitutes one Monte Carlo Step per Site (MCSS) denoted as unit of time ($t$) used in our simulation.

Instantaneous magnetisation is recorded at each MCSS,
\begin{equation}\label{m(t)}
   m(t)= \frac{1}{N_s}\sum_{i} S{_{i}}. 
\end{equation}

\section {\bf {Results:}}
In our present study we have thoroughly analysed the continuous (ferro to para) phase transition in the  
\textcolor{blue}{Ising ferromagnetic thick cubic shell} with $L=50$, focusing on how the critical point shifts with changes in shell thickness $\Delta$. An approximate estimation of $T_c$ is necessary to maintain the equivalent thermal condition across the systems with different shell thickness, ensuring that the thermal conditions remain comparable throughout the analysis. Transient behaviours for example, relaxation and reversal of magnetisation, are also investigated for Ising cubic shell with varying thickness $\Delta$.    
\subsection{Ferromagnetic Phase transition:}
In order to study the continuous ferro-para phase transition, the Ising ferromagnetic thick cubic shell with $L=50$ and thickness $\Delta$  is gradually cooled down in small temperature steps of $\delta T =0.05$, starting from high temperature random spin configuration (paramagnetic phase). At each time (MCSS) step, the corresponding thermodynamic quantities of the system are measured.
The instantaneous magnetisation $m(t)$ has been calculated. The equilibrium magnetisation in the absence of external field ($h_{\rm ext}=0$) is defined as $M=\langle m \rangle$, while susceptibility is obtained using $\chi = \frac{N_s}{k_BT}(\langle m^2 \rangle - {\langle m \rangle}^2)$. Here $\langle .. \rangle$ denotes the time average of thermodynamic quantities, approximately identical to the ensemble average in the ergodic limit. Thermodynamic quantities are recorded and averaged over $10^4$ MCSS to ensure more accurate and stable measurements. As an example, for a cubic shell with $L=50$ and $\Delta=5$, the total length of the simulation is 75000 MCSS. Out of which initial 65000 MCSS are discarded to ensure the ergodic limit and the statistical measurements are averaged over the rest $10^4$ MCSS. The time-averaged thermodynamic quantities are further averaged over 25-50 random samples depending on the system's size and thickness to enhance statistical accuracy.

\begin{figure}[h!tpb]
\begin{center}
  \includegraphics[angle=0, width=0.80\textwidth]{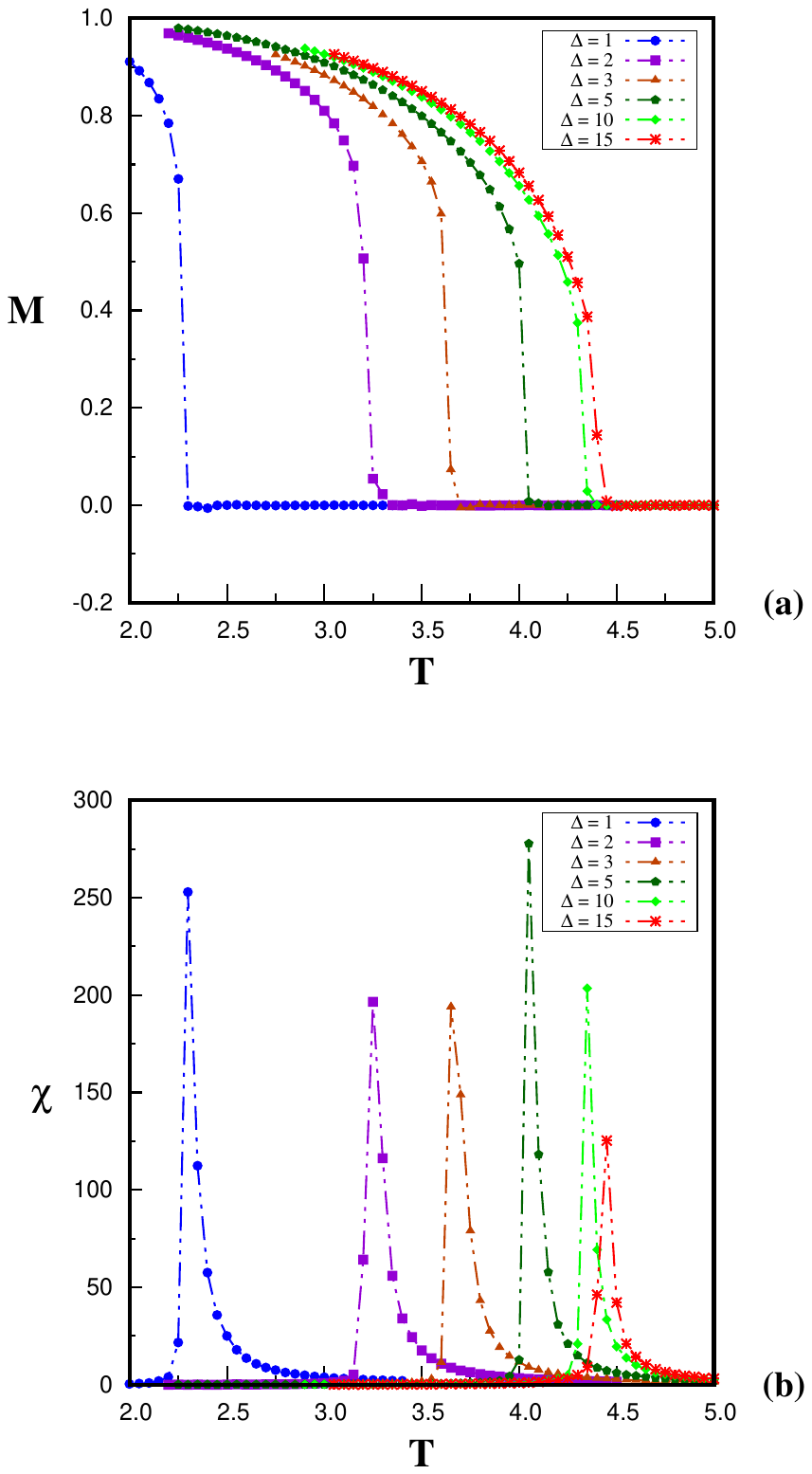}
  \caption{Thermodynamic quantities i.e., (a) equilibrium magnetisation $M$ and (b) susceptibility $\chi$ are studied as the functions of temperature ($T$) and demonstrated for Ising ferromagnetic cubic shell of size $L=50$ having different thickness ($\Delta$).}
  \label{fig:Criticalpoint}
\end{center} 
\end{figure}

We have  thoroughly observed the variation of magnetisation ($M$) and susceptibility ($\chi$) with temperature ($T$) and graphically demonstrated in the Figure \ref{fig:Criticalpoint} for different values of thickness $\Delta$ of the Ising cubic shell system. At high temperature, the order parameter i.e., magnetisation ($M$) of the system is almost zero, ensuring the paramagnetic phase. As we decrease the temperature, the magnetisation becomes non-zero at the transition temperature (or pseudo-critical temperature $T_c^p$). This is depicted in Figure \ref{fig:Criticalpoint}(a). The susceptibility of the system exhibits a sharp peak at the transition temperature ($T_c^p$). The position of the peak of the susceptibility, as shown in Figure \ref{fig:Criticalpoint}(b) provides an estimate of pseudo-critical temperature ($T_c^p$) of the Ising cubic shell for various thickness values $\Delta=$1, 2, 3, 5, 10, 15, 20, 22 and 25.

\begin{figure}
\centering
 \includegraphics[angle=0, width=1.00\textwidth]{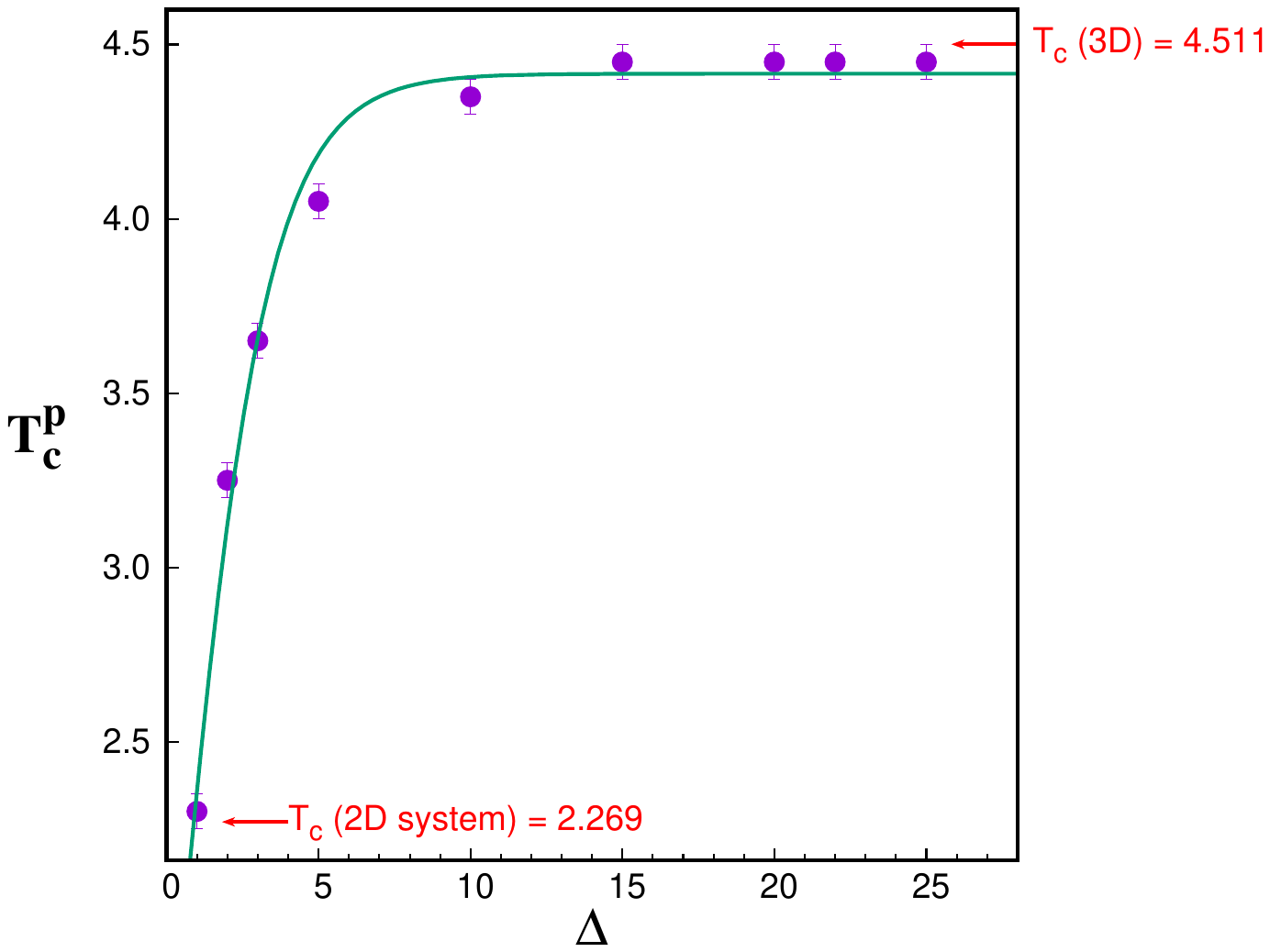}
 \caption{The variation of pseudo-critical temperature ($T_c^p(\Delta)$) with thickness $\Delta$ of the cubic shell $T_c^p(\Delta)$ and studied as a `$\tanh$' function of $\Delta$ i.e., $T_c^p(\Delta)=a\times \tanh(b~\Delta) + c$. The best fitted line is shown in solid green line. The fitted parameters : $a = 2.98 \pm 0.19 $, $b = 0.321 \pm 0.026$ and $c = 1.43 \pm 0.19 $. }
 \label{fig:Tc-delta}
\end{figure}

The transition temperature of an Ising ferromagnetic cubic shell with $L=50$ and thickness $\Delta=1$ is estimated to be $T_c^p=2.30 \pm 0.05$, remarkably close to Onsager's exact value of $T_c= 2.269$ (in $J/k_B$ unit) for a two-dimensional Ising system, which suggests that the ferro-para phase transition in the thinnest cubic shell (monolayer) closely resembles that of a flat two-dimensional system. Our results reveal that the transition temperature increases with the increase in the thickness of the cubic shell. This upward shift in $T_c^p$ continues up to a saturating value of $\Delta$ (around $\Delta=15$ here), indicating a direct dependence of pseudo-critical temperature on the shell-thickness before it stabilizes at higher values of $\Delta$. These observations lead us to the conclusion that beyond the saturating value of shell-thickness ($\Delta$ larger than 15) the presence of non-magnetic cavity within the thick shell of Ising spins does not influence the equilibrium ferro-para phase transition. Figure \ref{fig:Tc-delta} presents the pseudo-critical temperature ($T_c^p$) as a function of the cubic shell's thickness $\Delta$. As seen from Figure \ref{fig:Tc-delta}, the pseudo-critical temperature $T_c^p$ increases with thickness and after reaching a saturating value of $\Delta$, $T_c^p$ approaches Monte Carlo estimated value for 3D Ising ferromagnetic system $T_c=4.511~J/k_B$. The variation of $T_c^p$ with thickness $\Delta$ is well described by the function $T_c^p (\Delta) = a \times \tanh(b~\Delta) + c$, with best-fit parameters $a = 2.98 \pm 0.19 $, $b = 0.321 \pm 0.026$ and $c = 1.43 \pm 0.19 $. This fit, represented by the solid green line in Figure \ref{fig:Tc-delta} yields the value of statistical $\chi^2=0.046$ with degrees of freedom = 6. This finding is significant as it provides new insights into the equilibrium behaviour of Ising ferromagnetic thick cubic shells and has not been reported elsewhere.

\subsection{Magnetic Relaxation behaviour:}
The ferromagnetic system is initially perturbed from its equilibrium state by applying a very strong magnetic field, forcing it into a perfectly ordered configuration with magnetisation $m=1$. When the external perturbation (strong magnetic field) is abruptly removed, the system relaxes back to its steady state of equilibrium through the relaxation of magnetisation.

In this study, we have investigated the magnetic relaxation in the Ising ferromagnetic cubic shell. We have started from an initial state where all spins are aligned upwards (as if the system is in a strong magnetic field). This corresponds to a state of maximum magnetisation and certainly not the equilibrium state at any temperature in paramagnetic phase. Now, immediately after the removal of the strong field, the magnetisation of the system decays in time until it reaches equilibrium (with $m(t)$ vanishingly small) at any finite temperature $T$ (above $T_c^p$). The temperature of the system is maintained in the paramagnetic phase, specifically at $T=1.10~T_c^p(\Delta)$. 
\textcolor{blue}{It is worthmentioning that the condition $T=1.10~T_c^p(\Delta)$  serves as an equivalent thermal condition for all values of shell thickness ($\Delta$).}

\begin{figure}
\centering
 \includegraphics[angle=0, width=1.00\textwidth]{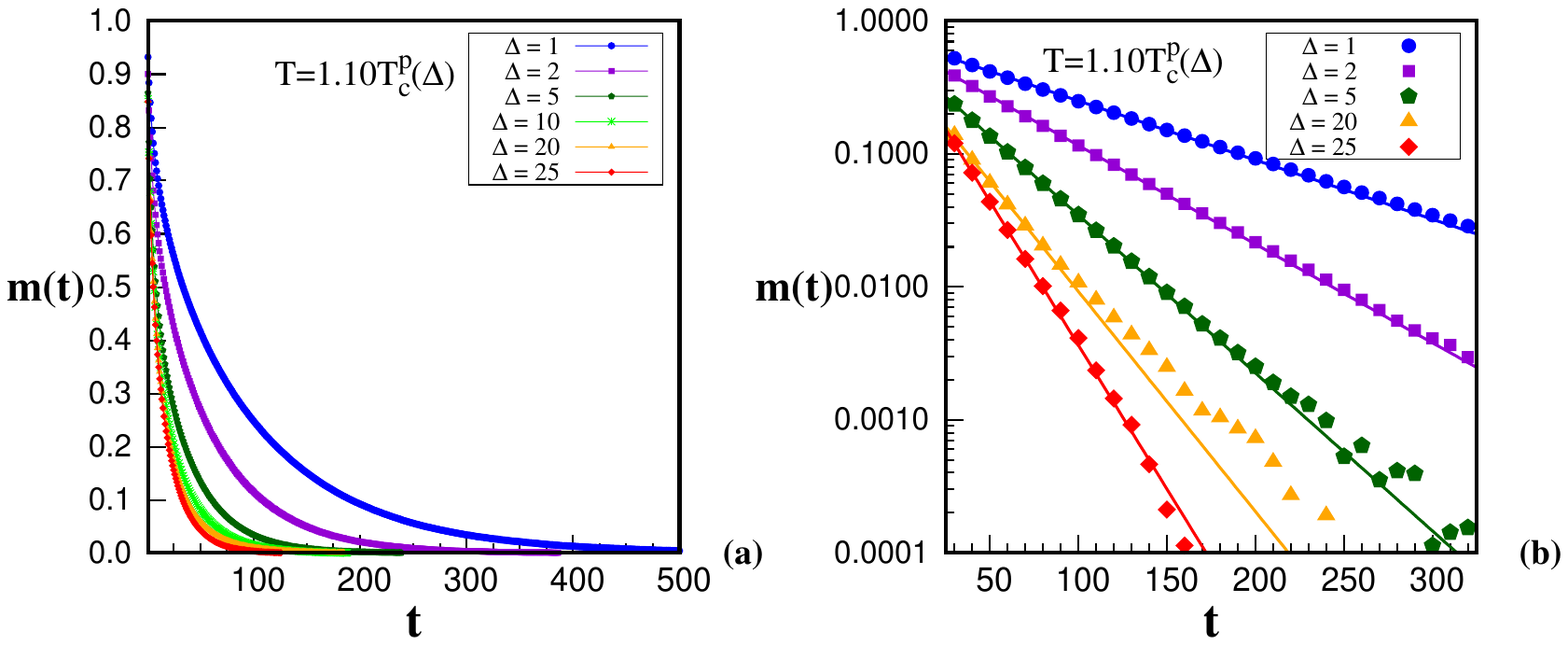}
 \caption{(a) The  decay of magnetisation $m(t)$ with time ($t$) in the Ising ferromagnetic thick cubic shell, where external field $h_{ext} = 0$ and temperature ($T$) of the system fixed at $T=1.10T_c^p(\Delta)$ (paramagnetic phase). Relaxation of magnetisation are plotted for the different values of shell-thickness $\Delta$. Results are obtained by averaging over 8000-25000 random samples.\\ (b) The semi-logarithmic plot of decay of magnetisation for selected values of shell-thickness. $\Delta=$ 1, 2, 5, 20 and 25. The data are fitted to a exponential function $f(x)=a\exp(-cx)$, represented by solid lines. Here $f(x)$ stands for $m(t)$ and $x$ represents $t$. The exponential nature of magnetic relaxation is evident (equation-4). It may be noted here that all
 data points (in (a)) are not shown in (b).}
 \label{fig:relaxation}
\end{figure}

\begin{figure}
\centering
 \includegraphics[angle=0, width=1.00\textwidth]{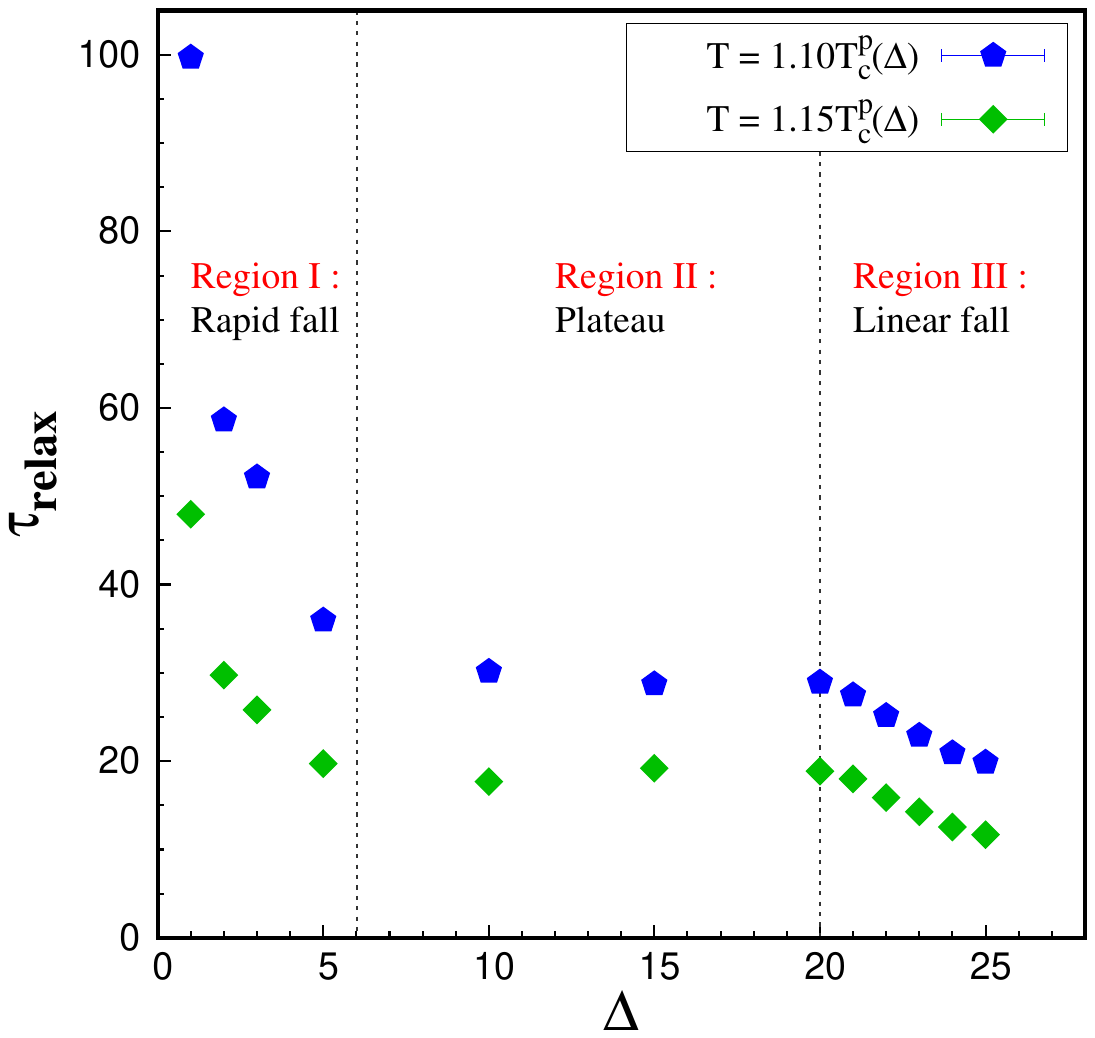}
 \caption{Dependence of Relaxation time $ \tau_{\rm relax}$ on the thickness of shell $\Delta$. The systems are kept at $T=f \times T_c^p (\Delta)$. We have considered  $f=1.10$ and $1.15$ for all values of the thickness($\Delta$) of the shell. Three distinct regimes are identified, namely rapid fall($\Delta=$ 1-6), plateau ($\Delta=$ 6-20) and linear region ($\Delta=$ 20-25).}
\label{fig:relaxtime}
\end{figure}

 Here, Figure \ref{fig:relaxation} illustrates the time evolution or the relaxation of magnetisation for different thickness $\Delta$ of the Ising ferromagnetic cubic shell of $L=50$, presented in the linear scale (\ref{fig:relaxation}(a)) and semi-log scale (\ref{fig:relaxation}(b)), respectively. The results are obtained, averaging over 8000-25000 random samples, depending on the system size to ensure statistical accuracy. The semi-logarithmic plot of decay of the magnetisation, as shown in Figure \ref{fig:relaxation}(b), clearly exhibits the exponential nature of magnetic relaxation of the following form, 
\begin{equation}
 m(t) \sim \exp(-t/\tau_{\rm relax})   
\end{equation}
\textcolor{blue}{where $\tau_{\rm relax}$ is the characteristic time associated with magnetic relaxation called {\it relaxation time}. The data of magnetisation decay $m(t)$ have been analysed by fitting to an exponential function $f(t)=a\exp(- c~t)$, where $a$ and 
$c(=b\times 10^{-3})$ are fitting parameters corresponding to the initial magnetization ($m(0)$) and decay rate, respectively.
It may be noted here for an exponential decay  $m(t)=m(0)\exp(-c~t)$, the logarithm of magnetization, $log~m(t)$, varies linearly with time $t$.
 The magnitude of the slope will be equal to $c$. One can easily identify that $\tau_{relax} = 1/c$ (comparing with equation-4).}
 The solid lines in Figure \ref{fig:relaxation}(b) represent the exponential fit for $m(t)$ on semi-logarithmic scale. The relaxation time $\tau_{\rm relax}$ is determined by taking the inverse of the decay rate ($c$) represented by the fitted parameter $c=b\times 10^{-3}$ of the function $f(t)$. The best-fit values of the fitting parameters are provided in Table \ref{tab:table1}.
\begin{table*}[h!]
  \caption {Best fitted parameters (Cubic shell) corresponding to Figure \ref{fig:relaxation} with $T=1.10T_c^p(\Delta)$.}
  \label{tab:table1}
  \begin{tabular*}{\textwidth}{@{\extracolsep{\fill}}llllll}
    \hline
    Thickness ($\Delta$) & $b$ & \textcolor{blue}{ $\tau_{relax}=1/c$} &   $\chi^2$ &  DOF  \\ 
    
    \hline 
     1 & $10.032 \pm 0.008$ & \textcolor{blue}{99.685} &0.00082 & 489 \\
     
     2 & $17.061 \pm 0.012$ & \textcolor{blue}{58.612} &0.00022 & 385  \\   
     
     5 & $27.823 \pm 0.042$ & \textcolor{blue}{35.941} &0.00023 & 229 \\
     
     20 & $34.571 \pm 0.139$ & \textcolor{blue}{28.926} &$4.83 \times 10^{-5}$ & 201\\
     
     25 & $50.319 \pm 0.083$ & \textcolor{blue}{19.873} &$1.94 \times 10^{-5}$ & 114 \\
     
    \hline
  \end{tabular*}
\end{table*}
\textcolor{blue}{The DOF (degrees of freedom) is defined as DOF$=N_{Total}-N_{Parameters}$, where $N_{Total}$ and $N_{Parameters}$ are the total number of data points and the total number of fitting parameters respectively. In the present case, the total number of fitting parameters is two.}
We have systematically recorded the relaxation time $\tau_{\rm relax}$ of the Ising ferromagnetic cubic shell by varying its shell-thickness $\Delta$ at two different temperatures, say $T=1.10T_c^p(\Delta)$ and $1.15T_c^p(\Delta)$. The dependence of $\tau_{\rm relax}$ on thickness $\Delta$ is depicted in Figure \ref{fig:relaxtime}. The obtained results reveal that the relaxation time decreases as we increase the shell-thickness. In the beginning, the relaxation time $\tau_{\rm relax}$ exhibits a rapid decrease with shell-thickness $\Delta$, suggesting that thinner shells experience more prolonged relaxation processes. This rapid fall region is extended within a range of values of thickness 
$\Delta$ (say $\Delta=1-6$). As $\Delta$ continues to increase, the relaxation time approaches a near-saturation point for intermediate values of $\Delta$ (say $\Delta=6-20$), indicating a regime where the influence of shell-thickness on the relaxation time is insignificant. In that plateau region, the relaxation dynamics become less sensitive to the existence of internal non-magnetic cavity. Beyond this intermediate regime ($\Delta > 20$), $\tau_{\rm relax}$ resumes a linear decrease with increasing $\Delta$, highlighting a renewed sensitivity of relaxation dynamics on the shell thickness at larger values. Our findings indicate a complex interplay between magnetic relaxation and shell-thickness. This is interestingly new result. This strange behaviour is not observed in the solid cube of Ising ferromagnet.

\subsection{Metastable behaviour:} 

\textcolor{blue}{Below the critical temperature, a ferromagnetic system exhibits metastable behaviour. When the system is initially prepared with all spins aligned upward and then allowed to evolve in the presence of a small magnetic field applied in the opposite direction, it remains in the metastable phase for a certain time duration. This duration  is referred to as the metastable lifetime. Eventually, the system achieves its stable state with negative magnetization.
The minimum time required by the system to achieve the state of negative magnetisation is defined as reversal time ($\tau_{\rm rev}$).}

\begin{figure}
    \centering
    \includegraphics[angle=0, width=1.00\textwidth]{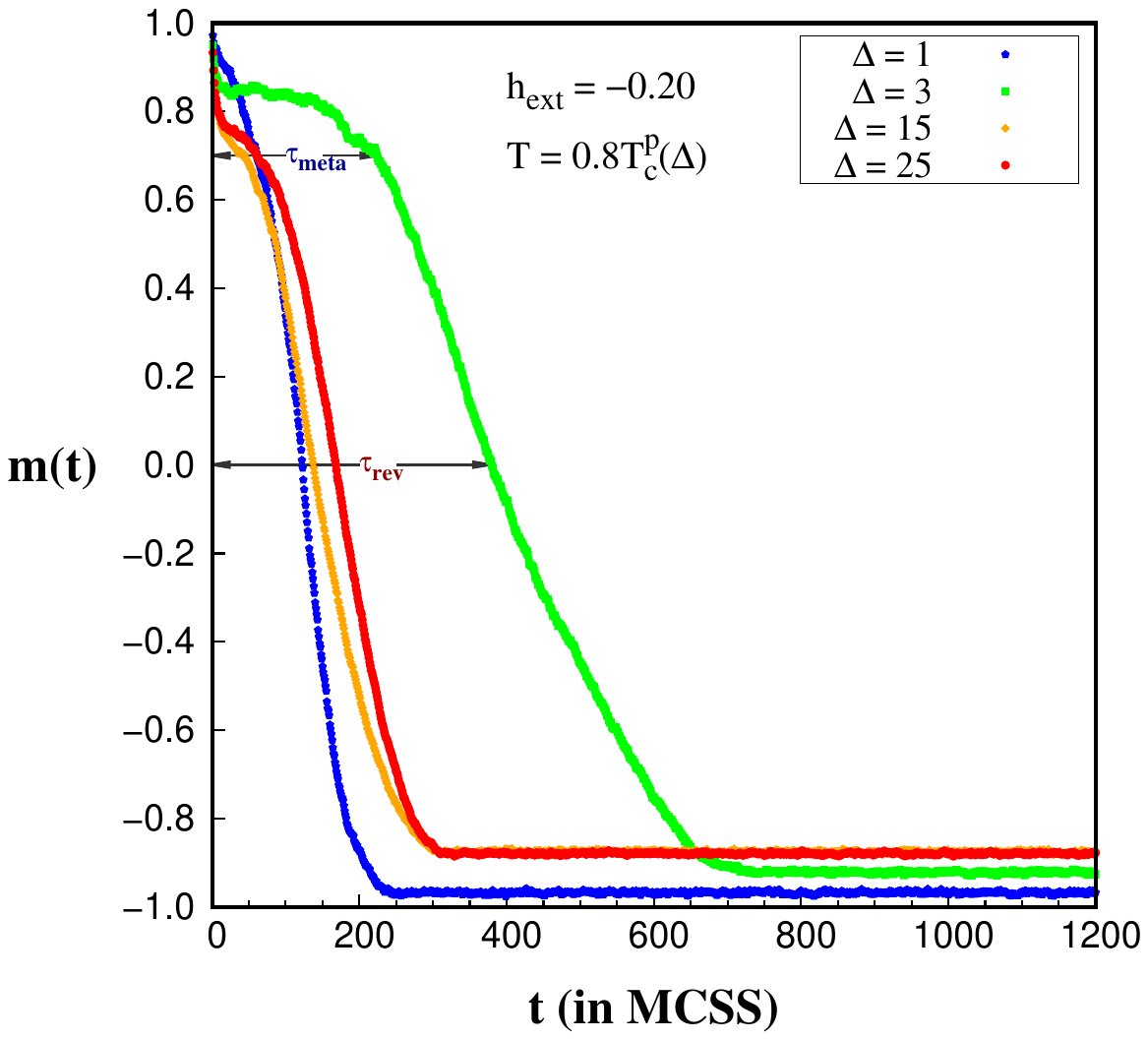}
    \caption{Decay of metastable state in the presence of external field $h_{ext}= -0.20$ in the Ising ferromagnetic cubic shell of $L=50$ having different values of thickness $\Delta =$ 1, 3, 15 and 25; Temperature maintained at $T = 0.8 T_c^p(\Delta)$, where $T_c^p(\Delta)$ is the pseudo-critical temperature of respective systems. The metastable lifetime $\tau_{\rm meta}$, defined as the minimum time steps required for the magnetisation to drop below $m=0.7$. The reversal time $\tau_{\rm rev}$ refers to the time by which the magnetisation crosses $m=0$ and changes sign.}
    \label{fig:decay}
\end{figure}

In this section, we provide a comprehensive analysis of the magnetisation reversal in the Ising ferromagnetic thick cubic shell, focusing on the effects of shell-thickness on the metastable behaviour. The Ising ferromagnetic shell of $L=50$, initially configured with all spins up i.e., $S_i = +1,~\forall i$ condition, has been exposed to a weak negative magnetic field $h_{\rm ext}= -0.20$. Temperature of the system is fixed at $T=0.8T_c^p(\Delta)$ (ferromagnetic phase). This choice is arbitrary and serves as an equivalent thermal condition for cubic shells for various shell thicknesses. We have critically examined the decay of metastable state for a single sample and illustrated in Figure \ref{fig:decay}, showing the system's metastable behaviour for various  shell thickness $\Delta=$ 1, 3, 15 and 25. These findings offer valuable insight into how the thickness of the shell influences the lifetime of the metastable state. 

In our analysis, we have referred to the metastable lifetime $\tau_{\rm meta}$ as the minimum time steps required for the magnetisation to drop below $m=0.7$. This method of defining the metastable lifetime is used in previous studies \cite{acharyya}. Similarly, the reversal time\cite{mapre} denoted as $\tau_{\rm rev}$ refers to the time by which the magnetisation crosses $m=0$ and changes sign. To obtain statistically significant data, we have calculated the mean value of $\tau_{\rm rev}$ as well as $\tau_{\rm meta}$ by averaging over 1000-5000 random ferromagnetic samples. 

\begin{figure}[h!tpb]
    \centering
    \includegraphics[angle=0, width=0.95\textwidth]{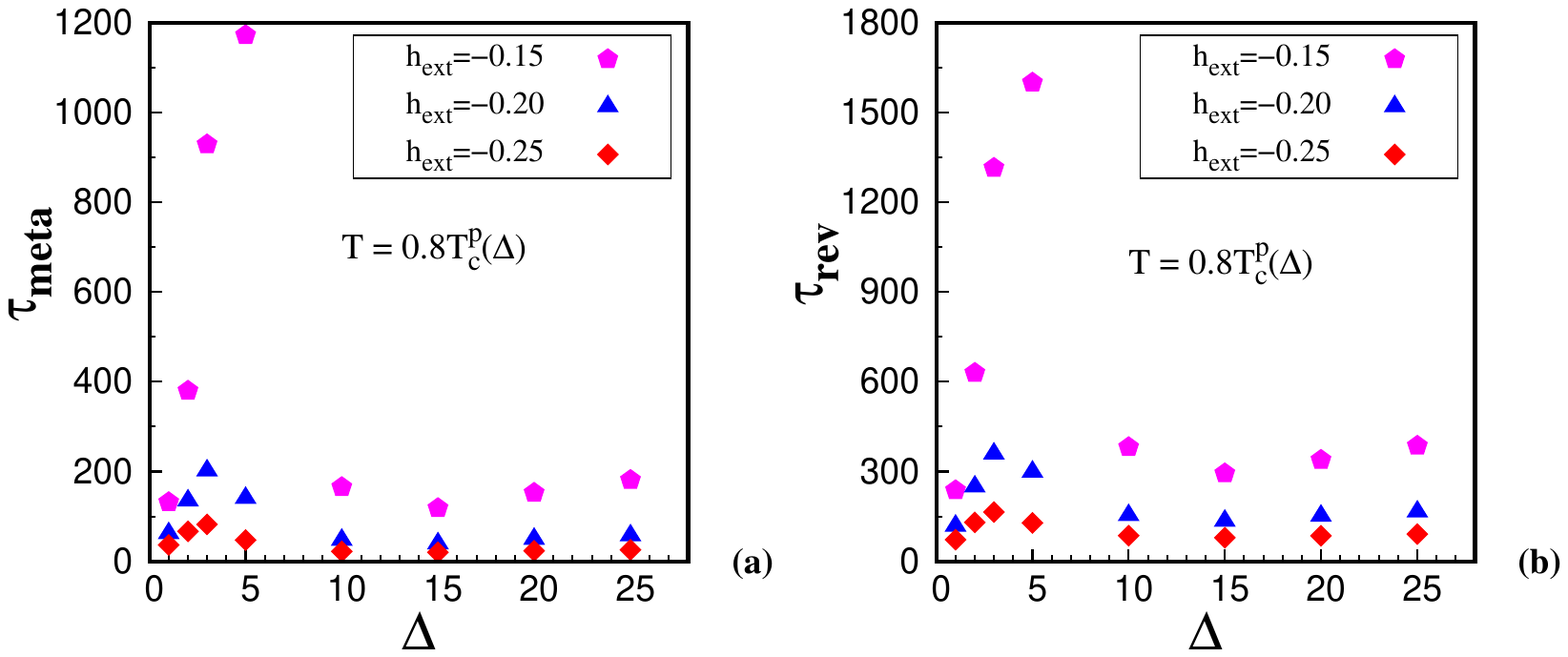}
    \caption{(a) $\tau_{\rm meta}$, the mean metastable lifetime non-monotonically varies with thickness $\Delta$ of the Ising ferromagnetic shell.
    (b) Reversal time $\tau_{\rm rev}$ shows similar non-monotonic variation with $\Delta$. Both results provided for three different strengths of external field $h_{\rm ext}=$ -0.15, -0.20 and -0.25. Temperature maintained at $T = 0.8 T_c^p(\Delta)$. Data averaged over 1000-5000 samples.}
    \label{fig:revtime}
\end{figure}

The variation of mean $\tau_{\rm meta}$ with thickness $\Delta$ is graphically represented in Figure \ref{fig:revtime} (a) for three distinct external field strengths $h_{\rm ext}= -0.15,~-0.20 ~\rm and~-0.25$. Our findings reveal that the mean metastable lifetime $\tau_{\rm meta}$ changes non-monotonically with increasing shell thickness $\Delta$. Furthermore, we have studied the mean reversal time $\tau_{\rm rev}$ as a function of thickness $\Delta$ for \textcolor{blue}{a fixed value} of the field strength, as shown in Figure \ref{fig:revtime} (b). Both of these two time scales $\tau_{\rm meta}$ and $\tau_{\rm rev}$ exhibit almost a similar kind of non-monotonic variation with $\Delta$. The mean reversal time $\tau_{\rm rev}$ is observed to be at its minimum for the thinnest shell ($\Delta = 1$), indicating a relatively faster reversal process. Interestingly, a specific range of shell-thickness ($\Delta=3-5$) \textcolor{blue}{has been identified that maximizes both time scales} ($\tau_{\rm meta}$ and $\tau_{\rm rev}$). A particular thickness within this range leads to significantly prolonged metastability. Notably, no similar results have been reported elsewhere and this results are remarkably different from that observed in solid cube of Ising ferromagnet.

\section{Summary:} 
In this manuscript, we have reported the results of both equilibrium and nonequilibrium behaviours of a thick cubic Ising ferromagnetic shell, \textcolor{blue}{investigated through Monte Carlo simulations.} The results are obtained using Monte Carlo simulation. In the equilibrium case, the ferro-para phase transition is studied. The transition temperature (pseudo-critical temperature $T_c^p$) is obtained from the temperature which maximizes the susceptibility. The pseudo-critical temperature has been found to depend remarably on the thickness ($\Delta$) of the cubic shell. 
Our numerical results show that the pseudo-critical temperature increases as the thickness of the cubic shell increases. For the thinnest shell (monolayer) the pseudo-critical temperature is close to the transition temperature of a two-dimensional Ising ferromagnet\cite{ishita2}. \textcolor{blue}{Conversely, in the limit of maximum thickness, where the cubic shell becomes a solid cube, } the pseudo-critical temperature is approximately equal to the Monte Carlo estimate of the ferro-para transition temperature of a solid cube. These results are quite obvious. In the intermediate region, the variation of the pseudo-critical temperature with the thickness of the shell \textcolor{blue}{is well demonstrated by the fitting function } like $T_c^p(\Delta)=a\times \tanh(b~\Delta) + c$, where the values of $a$, $b$ and $c$ are $a = 2.98 \pm 0.19 $, $b = 0.321 \pm 0.026$ and $c = 1.43 \pm 0.19 $ estimated for the best fit. \textcolor{blue}{It is worth noting that our estimated} critical temperatures for $\Delta=2$ and $\Delta=3$, \textcolor{blue}{closely match those predicted for the } layered Ising ferromagnet in mean field and interfacial approximations\cite{lipowski}. This is an interesting result of course and has not been found before in the literature. 

In the nonequilibrium case, \textcolor{blue}{we have investigated two key aspects of the thick cubic Ising shell: (i) the relaxation behaviour and (ii) the metastable behaviour.} In the relaxation behaviour, we have allowed the system to relax from a state where all spins are up. The state of all spins are up is the equilibrium state (ground state) for $T=0$. 
\textcolor{blue}{ At any finite temperature, however, the system evolves toward its true equilibrium state via the relaxation of magnetisation
.} The equivalent thermal condition is prepared as $T= f \times T_c^P(\Delta)$. We have used $f=1.10~\rm and~1.15$ for all values of the thickness ($\Delta$) of the shell. \textcolor{blue}{Under these conditions, the system exhibits exponential relaxation for all values of $\Delta$.} However, as the shell-thickness ($\Delta$) increases, the relaxation time $\tau_{\rm relax}$ is found to decrease. It may be noted here that for smaller values of thickness the relaxation time decreases rapidly, in an intermediate range of the shell-thickness we have observed a plateau region where the relaxation time is almost independent of shell-thickness. For larger values of shell-thickness the relaxation time linearly decreases with the thickness. As far as we are aware, we have not encountered similar studies or results in the literature.

\textcolor{blue}{The decay of metastable state is another remarkable feature of the non-equilibrium behaviour. The Ising ferromagnet shows metastability. In this case, the well studied\cite{acharyya} fact is that 
the metastable lifetime depends on the magnitude of the applied magnetic field.} In this paper, we have studied the dependences of metastable lifetime on the thickness of the thick cubic shell of Ising ferromagnet. Interestingly, the metastable lifetime also shows a non-monotonic variation with the thickness of the cubic shell. Initially, as the value of shell-thickness increases, the metastable lifetime has been found to increase. After a specific value of the shell-thickness the metastable lifetime has
been found decrease. As a result, there exists a specific value of the shell-thickness, for which the metastable lifetime becomes maximum. These results are not found in any previous study.

Our study may trigger the interest of the experimentalist to investigate these kinds of behaviours in real ferromagnetic sample having cavity. Depending on the size of the cavity (non-magnetic), the transition temperature and the metastable lifetime can be regulated by varying the size of the cavity inside the ferromagnetic sample.

\vskip 1cm

\noindent {\bf Acknowledgements:}  
IT acknowledges UGC JRF, Govt. of India for financial support. We are thankful to the computational facilities provided by Presidency University, Kolkata. 
\vskip 0.2cm
\noindent {\bf Data availability statement:} Data will be available on reasonable request to Ishita Tikader.

\vskip 0.2cm

\noindent {\bf Conflict of interest statement:} We declare that this manuscript is free from any conflict of interest.
\vskip 0.2cm

\noindent {\bf Funding statement:} No funding was received particularly to support this work.

\vskip 0.2cm

\noindent {\bf Authors’ contributions:} Ishita Tikader- Developed the code, prepared the figures, and wrote the manuscript.
Muktish Acharyya-Conceptualized the problem, analysed the results and wrote the manuscript.


\newpage

\newpage

\newpage
\newpage
\newpage
\newpage
\newpage

\end{document}